# STUDY ON THE MAGNETIC MEASUREMENT RESULTS OF THE INJECTION SYSTEM FOR CSNS/RCS*

M.Y. Huang[1,2#], S.N. Fu[1,2], N. Huang[1], L.H. Huo[1], H.F. Ji[1,2], W. Kang[1,2], Y.Q. Liu[1,2], J. Peng[1,2], J. Qiu[1,2], L. Shen[1,2], S. Wang[1,2], X. Wu[1,2], S.Y. Xu[1,2], J. Zhang[1,2], G.Z. Zhou[1,2]

1：Institute of High Energy Physics, Chinese Academy of Sciences, Beijing, China
2：Dongguan Institute of Neutron Science, Dongguan, China

*Abstract*

A combination of the H⁻ stripping and phase space painting method is used to accumulate a high intensity beam in the Rapid Cycling Synchrotron (RCS) of the China Spallation Neutron Source (CSNS). The injection system for CSNS/RCS consists of three kinds of magnets: four direct current magnets (BC1-BC4), eight alternating current magnets (BH1-BH4 and BV1-BV4), two septum magnets (ISEP1 and ISEP2). In this paper, the magnetic measurements of the injection system were introduced and the data analysis was processed. The field uniformity and magnetizing curves of these magnets were given, and then the magnetizing fitting equations were obtained.

## INTRODUCTION

The China Spallation Neutron Source (CSNS) is a high power proton accelerator-based facility [1]. It consists of an 80 MeV H⁻ linac (upgradable to 250 MeV for CSNS-II), a 1.6 GeV Rapid Cycling Synchrotron (RCS), a solid tungsten target station, and some instruments for neutron applications [2]. With a repetition rate of 25 Hz, the RCS, which accumulates an 80 MeV injection beam, accelerates the beam to the designed energy of 1.6 GeV and extracts the high energy beam to the target. The design goal of beam power for CSNS is 100 kW and can be upgradable to 500 kW [3].

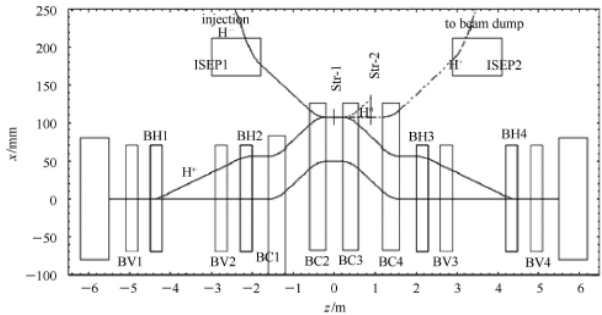

Figure 1: Layout of the RCS injection system.

For the high intensity proton accelerators, the injection with H⁻ stripping and phase space painting is actually a practical method [4] which is used for CSNS/RCS. Figure 1 shows the layout of the RCS injection system. It consists of a fixed horizontal bump (four direct current (DC) dipole magnets, BC1-BC4), a horizontal painting bump (four alternating current (AC) dipole magnets, BH1-BH4), a vertical painting bump (four AC dipole magnets, BV1-BV4), two septum magnets (ISEP1, ISEP2), and two stripping foils (Str-1, Str-2) [5].

In order to obtain the field uniformity and magnetizing curves of different kinds of magnets in the injection system for CSNS/RCS, the magnetic measurements need to be done. By using some codes of numerical analysis, the measurement results can be processed. Then the field uniformity and magnetizing curves can be given and the magnetizing equations can be fitted.

## MEASUREMENTS OF BC MAGNETS

For the injection system of CSNS/RCS, four DC dipole magnets, BC1-BC4, give a fixed horizontal bump in the middle for an additional closed-orbit shift of 60 mm. The physics design parameters of the four BC magnets are the same and they share only one power supply.

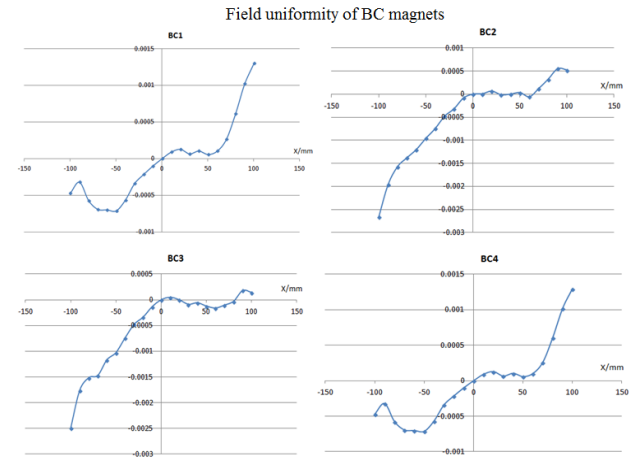

Figure 2: Field uniformity of the four BC magnets.

After the measurements of the four BC magnets, their field uniformity can be obtained, as shown in Fig. 2. It can be known that the field uniformity of any one BC magnet is smaller than ±0.3% which meets the physics design requirement (<±0.5%). In addition, it can be found that there are some differences between the four BC magnets although they are the same in the physics design.

The magnetizing curves of the four BC magnets and their auxiliary coils also can be given after the magnetic measurements. By using some codes of numerical analysis, the magnetizing curves can be fitted. After the analysis of fitting errors, it can be known that the linear of

___

*Work supported by National Natural Science Foundation of China (Project Nos. 11205185)

#huangmy@ihep.ac.cn

the magnetizing curves is very good. Then the linear fitting equations of the four BC magnets and their auxiliary coils can be given, as shown in Table 1.

Table 1: The Magnetizing Fitting Equations of the Four BC Magnets and their Auxiliary Coil

| Magnet | BC | Equation (BL/T·mm, I/A) |
|---|---|---|
| Main | BC1 | BL = 0.1821×I＋0.06024 |
|  | BC2 | BL = 0.1822×I＋0.1789 |
|  | BC3 | BL = 0.1823×I＋0.182 |
|  | BC4 | BL = 0.1824×I＋0.1789 |
| Auxiliary coil | BC1 | BL = 0.08891×I＋0.06674 |
|  | BC2 | BL = 0.08899×I＋0.007756 |
|  | BC3 | BL = 0.08971×I＋0.02102 |
|  | BC4 | BL = 0.08911×I＋0.01571 |

## MEASUREMENTS OF BH MAGNETS

For CSNS/RCS, the transverse phase space painting method is used for injecting a small emittance beam from the linac into the large ring acceptance. Four AC dipole magnets, BH1-BH4, are used for horizontal painting and another four AC dipole magnets, BV1-BV2, are used for vertical painting.

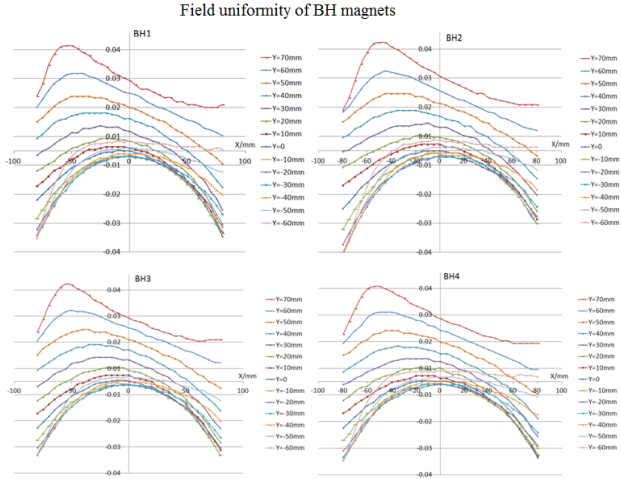

Figure 3: Field uniformity of the four BH magnets.

The physics design parameters of the four BH magnets are the same and they share only one power supply. The design requirement of the field uniformity for BH magnets is smaller than ±1.5%. After the magnetic measurements, the field uniformity of the four BH magnets can be obtained, as shown in Fig. 3. It can be found that, in some areas, the field uniformity of BH magnets is larger than ±1.5% which doesn't meet the physics design requirement. After multi-turn tracking simulation by the code ORBIT [6], the effects of bad field uniformity for the real particle distribution area can be studied [7] and it can be found that the anti-correlated painting method is more suitable.

As shown in Fig. 4 (left), the current curves of the power supply of the four BH magnets are given, and then the measurements of magnetizing curves can be done. In Fig. 4 (right), the magnetizing curves of BH magnets are given. It can be known that the rising rate and falling rate are very important factors to impact the magnetizing curves. In addition, the error becomes lager while the current smaller than 3000 A. Furthermore, by using some codes of numerical analysis, the magnetizing curves can be fitted and the fitting equations can be obtained.

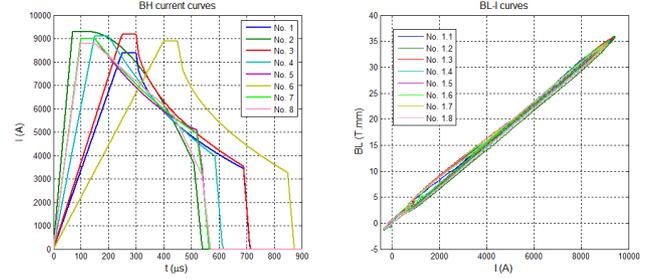

Figure 4: The current curves (left) and magnetizing curves (right) of BH magnets.

## MEASUREMENTS OF BV MAGNETS

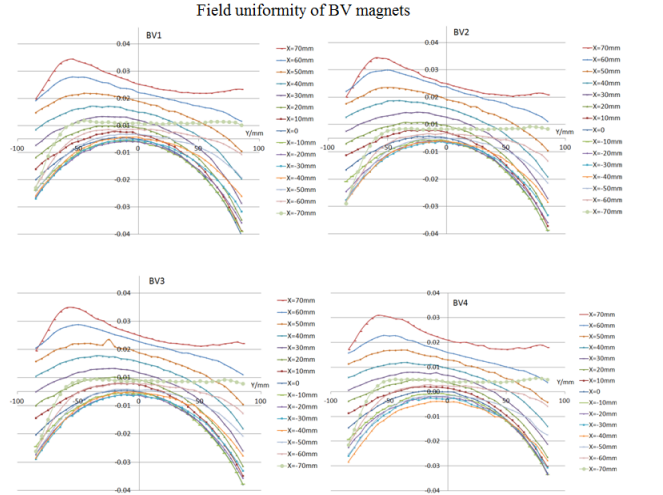

Figure 5: Field uniformity of the four BV magnets.

Similar to the four BH magnets, the physics design parameters of the four BV magnets are the same and they also share only one power supply. The design requirement of the field uniformity for BV magnets is smaller than ±1.5%. After the magnetic measurements, the field uniformity of the four BV magnets can be obtained, as shown in Fig. 5. It can be found that, in some areas, the field uniformity of BV magnets can achieve about ±4% which is larger than ±1.5% and doesn't meet the physics design requirement. After multi-turn tracking simulation by the code ORBIT, the effects of bad field uniformity for the real particle distribution area can be studied and it can be found that the effects can impact the injection painting process.

The current curves of the power supply of the four BV

magnets are given in Fig. 6 (left), and the measurements of magnetizing curves can be done. Figure 6 (right) shows the magnetizing curves of BV magnets. It can be known that the rising rate and falling rate are very important factors to impact the magnetizing curves. In addition, the error becomes larger while the current is small. Furthermore, by using some codes of numerical analysis, the magnetizing curves can be fitted and the fitting equations can be obtained.

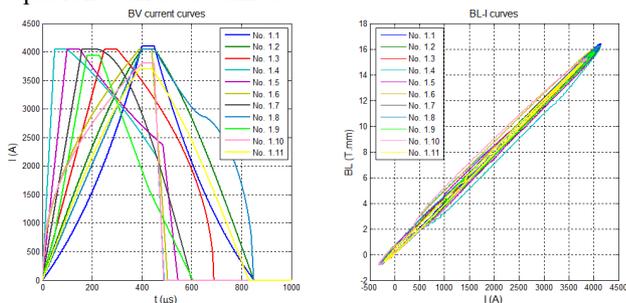

Figure 6: The current curves (left) and magnetizing curves (right) of BV magnets.

## MEASUREMENTS OF SEPTUM MAGNETS

There are two septum magnets in the injection system. The physics design parameters of the two septum magnets are the same, but they don't share one power supply. After the magnetic measurements, the field uniformity of the two septum magnets can be obtained, as shown in Fig. 7. It can be known that the field uniformity of any one septum magnet is smaller than ±0.4% which meets the physics design requirement (≤±0.5%).

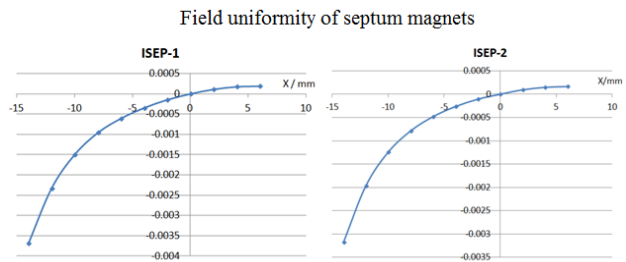

Figure 7: Field uniformity of the two septum magnets.

Table 2: The Magnetizing Fitting Equations of the Two Septum Magnets

| Septum | Equation (BL/T·mm, I/A) |
|---|---|
| ISEP-1 | BL = 0.1533×I＋0.1647 |
| ISEP-2 | BL = 0.1531×I＋0.2709 |

During the magnetic measurements, the magnetizing curves of the two septum magnets can be obtained. By using some codes of numerical analysis, the magnetizing curves can be fitted. Through the analysis of fitting errors, it can be known that the linear of the magnetizing curves is very good. Then the linear fitting equations of the two septum magnets can be given, as shown in Table 2.

## CONCLUSIONS

In this paper, the measurements of different kinds of magnets in the injection system were introduced. The field uniformity and magnetizing curves of different magnets were given. It can be found that the field uniformity of BC magnets and septum magnets meets the physics design requirement. However, in some areas, the field uniformity of BH and BV magnets cannot meet the physics design requirement. By using some codes of numerical analysis, the magnetizing curves of different kinds of magnets can be fitted, and the magnetizing fitting equations were obtained.

## ACKNOWLEDGMENT

The authors want to thank other CSNS colleagues for the discussions and consultations.